# Probabilistic Binary-Mask Cocktail-Party Source Separation in a Convolutional Deep Neural Network

Andrew J.R. Simpson [#1]

[#] *Centre for Vision, Speech and Signal Processing, University of Surrey*
*Surrey, UK*
[1] `Andrew.Simpson@Surrey.ac.uk`

*Abstract*—**Separation of competing speech is a key challenge in signal processing and a feat routinely performed by the human auditory brain. A long standing benchmark of the spectrogram approach to source separation is known as the ideal binary mask. Here, we train a convolutional deep neural network, on a two-speaker cocktail party problem, to make probabilistic predictions about binary masks. Our results approach ideal binary mask performance, illustrating that relatively simple deep neural networks are capable of robust binary mask prediction. We also illustrate the trade-off between prediction statistics and separation quality.**

## I. Introduction

Much work in source separation has focused on the so-called cocktail party problem [1], where a listener must selectively attend to speech within a background of competing speech noise. Study of this problem is partly motivated by the fact that the early auditory brain appears capable of facilitating this selective attention by maintaining separate spectro-temporal representations for concurrent streams of speech [2], [3].

A common approach to speech separation is to transform the mixture audio into a spectrogram representation and then to assign each time-frequency element (of the mixture spectrogram) to a particular source [4], [5]. Given spectrograms for each of the component source signals of the mixture, each time-frequency element of the mixture spectrogram may be attributed to the source with the largest magnitude in the source spectrogram. Thus, an 'ideal' separation may be derived – the ideal binary mask.

The ideal binary mask relies on the fact that concurrent speech features relatively little overlap in time-frequency space and serves as common benchmark in spectrogram-based source separation [4], [5]. By making full use of the spectrograms computed from the original sources, the ideal binary mask also defines the approximate performance ceiling of the approach. Thus, the key challenge in binary mask source separation is estimation of binary masks that are as similar as possible to the ideal masks estimated from the known sources.

In this paper, we employed a convolutional deep neural network (DNN) to learn the ideal binary mask for a two-speaker speech separation problem where speech from the two speakers is used as training data and where the model is then tested on new speech from the same speakers. From the model we obtained a probabilistic estimate of the ideal binary mask. Using objective source separation quality metrics, we analysed the performance of the model under different conditions of interpretation of the probabilistic binary mask predictions. Our results demonstrate that a convolutional DNN can approach the limits of the ideal binary mask and that interpretation can be optimized for different separation quality measures.

## II. Method

We consider a typical simulated cocktail party listening scenario featuring one male and one female voice speaking concurrently. The speakers were recorded separately (in mono), each reading from different stories. The two speech signals were adjusted to be of equal average intensity. To produce a mixture representative of a monaural cocktail party listening scenario, the monaural speech from each speaker was linearly summed to produce a monaural mixture signal (see Fig. 1).

The mixture and original speech signals were decimated to a sample rate of 4 kHz. The original speech signals and mixture were transformed into spectrograms using the short-time Fourier transform (STFT) with window size of 128 samples, overlap interval of 1 sample and a Hanning window. This provided spectrograms with 65 frequency bins. The phase component of each spectrogram was removed and retained for later use in inversion. From the source spectrograms a binary mask was computed where each element of the mask was determined by comparing the magnitudes of the corresponding elements of the source spectrograms and assigning the mask a '1' when the male voice had greater magnitude and '0' otherwise.

The magnitude-only mixture spectrogram computed from the first 2 minutes of the mixture signal and the respective ideal binary mask were used as training data. A subsequent 10 seconds of mixture spectrogram was held back for later use in testing the separation of the model. Note, phase was not used in training the model.

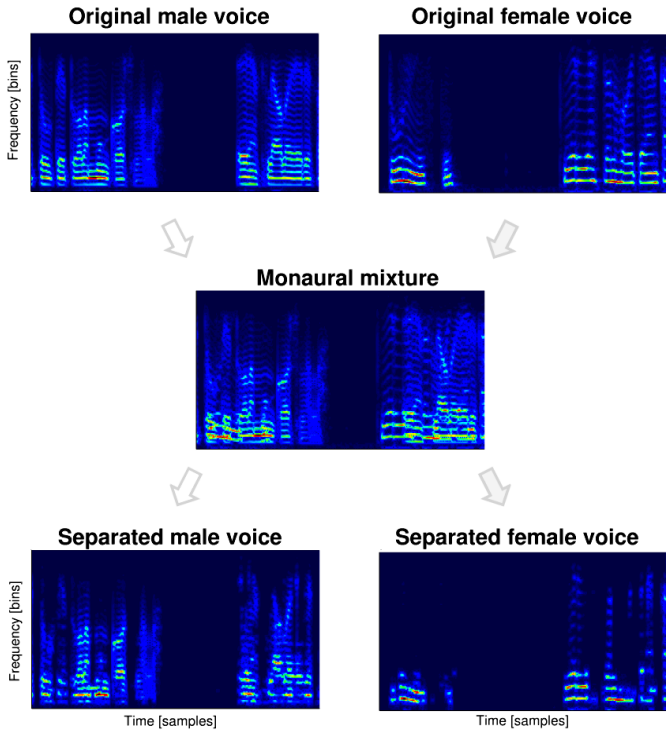

**Fig. 1. Monaural cocktail party source separation using a probabilistic convolutional deep neural network.** The upper pair of spectrograms plot a ~3-second excerpt from the original monaural audio for the male and female voice respectively. The middle spectrogram plots monaural mixture. The lower pair of spectrograms plot the respective separated channels ($\alpha = 0.99$). This excerpt features (coincidentally) approximately simultaneous utterances which result in overlapping vocalizations from both speakers in the mixture.

For training data, the mixture spectrogram and the corresponding input spectrograms were cut up into corresponding windows of 20 samples (corresponding to a quarter of a second). The windows overlaped at intervals of 10 samples. Thus, for every quarter-second window, for training the model there was a mixture spectrogram matrix of size 65x20 samples and an ideal binary mask matrix of the same size. This gave approximately 50,000 training examples. For the testing stage, 10 seconds of speech mixture was used at overlap intervals of 1 sample, giving approximately 40,000 test frames (which would ultimately be applied in an overlaping convolutional output stage). Prior to windowing, all spectrogram data was normalized to unit scale.

We used a feed-forward DNN of size 1300x1300x1300 units (65 x 20 = 1300). Each spectrogram window of size 65 x 20 was unpacked into a vector of length 1300. The DNN was configured such that the input layer was the mixture spectrogram (1300 samples). The DNN was trained to synthesize the ideal binary mask at its output layer. The DNN employed the biased-sigmoid activation function [6] throughout with zero bias for the output layer. The DNN was trained using 600 full iterations of stochastic gradient descent (SGD). Each iteration of SGD featured a full sweep of the training data. Dropout was not used in training. After training, the model was used as a feed-forward probabilistic device.

*Probabilistic Binary Mask.* In the testing stage, there was an overlap interval of 1 sample. This means that the test data described the mixture spectrogram in terms of a sliding window and the output of the model described predictions of the ideal binary mask in the same sliding window format. The output layer was sigmoidal and hence we may interpret these predictions in terms of the logistic function. Therefore, because of the sliding window, this procedure resulted in a distribution (size 20) of predictions for each time-frequency element of the mixture spectrogram. We chose to summarize this distribution by taking the mean and we evaluate the result in terms of an empirical confidence estimate, separately for each source, as follows: For each time-frequency element, of each source, we computed the mean prediction and applied a confidence threshold ($\alpha$);

$$M^A_{t,f} = \begin{cases} 1 & for \quad \frac{1}{T}\sum_{i=0}^{T} S_{t+i,f} > \alpha \\ 0 & for \quad \frac{1}{T}\sum_{i=0}^{T} S_{t+i,f} \leq \alpha \end{cases} \quad (1)$$

where $M^A$ refers to the binary mask for the male speaker, $T$ refers to the window size (20), $t$ is the time index, $i$ is the window index and $f$ is the frequency (bin) index into the mixture spectrogram ($S$). The corresponding (but independent) binary mask for the female speaker ($M^B$) is computed as follows;

$$M^B_{t,f} = \begin{cases} 1 & for \quad \frac{1}{T}\sum_{i=0}^{T} S_{t+i,f} < (1-\alpha) \\ 0 & for \quad \frac{1}{T}\sum_{i=0}^{T} S_{t+i,f} \geq (1-\alpha) \end{cases} \quad (2)$$

Thus, by adjustment of $\alpha$, masks at different levels of confidence could be constructed for both sources.

The respective masks were resolved by multiplication with the original (i.e., complex) mixture spectrogram and the resulting masked spectrograms were inverted with a standard overlap-and-add procedure. Separation quality (for the test data) was measured using the BSS-EVAL toolbox [7] and is quantified in terms of signal-to-distortion ratio (SDR), signal-to-artefact ratio (SAR) and signal-to-interference ratio (SIR). Separation quality was assessed at different confidence levels by setting different values of $\alpha$ in Eqs. 1 and 2.

## III. RESULTS

Fig. 1 plots spectrograms illustrating the stages of mixture and separation for a brief excerpt (~3 seconds) from the test data. The spectrograms for the source speech signals are shown at the top. The middle panel plots the mixture spectrogram and illustrates overlap of the two competing speech signals in time-frequency space. At the bottom of Fig. 1 are plotted the separated and re-synthesized audio for the male and female voices respectively ($\alpha = 0.99$). The separated spectrograms are reasonably faithful to the original sources but feature some points at which there are discontinuities in the partials, some missing energy and other small distortions.

Fig. 2 plots the various objective source separation quality metrics (SDR/SIR/SAR), computed over the entire 10-second test data, as a function of confidence level ($\alpha$) between the range of 0.001 and 0.999. SIR monotonically increases with the value of $\alpha$, SAR monotonically decreases with the value of $\alpha$, and SDR features a non-monotonic function which is a product of the crossover of SIR and SAR. Separation (SIR) and distortion (SDR) are at their lowest (poorest) for small values of $\alpha$, but the corresponding measure of artefacts (SAR) peak at the smallest value of $\alpha$. This is because at small values of $\alpha$ the masks resolve towards one (hence little separation occurs and hence few artefacts are introduced).

For reference, the equivalent measures for the ideal binary masks (computed directly from the source spectrograms) are SDR: 11.3, SIR: 21.6, SAR: 11.7. The closest performance from the model occurs at $\alpha = $ ~0.99, where we see an SIR value of ~21 (with corresponding SDR: ~6, SAR: ~6). Therefore, at large values of $\alpha$, our convolutional DNN performs at close to ideal binary mask levels.

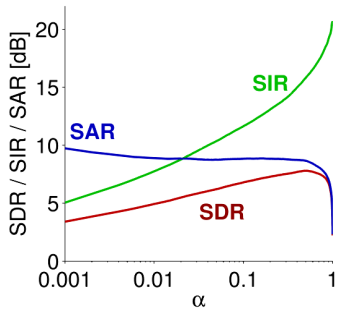

**Fig. 2. Separation quality as a function of $\alpha$.** Mean signal-to-distortion ratio (SDR, red), signal-to-interference (SIR, green), signal-to-artefact ration (SAR, blue), computed from the 10-second test audio using the BSS-EVAL toolkit [7]. Means are computed across both male and female sources. For reference, the ideal binary mask (computed from the actual source spectrograms for the male and female speech which were mixed to produce the test mixture) provides the following mean separation quality; SDR: 11.3, SIR: 21.6, SAR: 11.7.

## IV. DISCUSSION AND CONCLUSION

We have demonstrated that a convolutional deep neural network is capable of approaching the ideal binary mask separation performance. Our convolutional DNN is relatively simple and small scale, and was trained with relatively little data; only two minutes of training audio was provided for each voice. We have also demonstrated that the objective separation quality measures are dependent upon the probabilistic interpretation of the convolutional predictions made by the model. In particular, we have demonstrated that the model may in principle be optimized for either sound quality or separation/suppression and that these goals are mutually exclusive. However, our results also suggest that there is a comfortable global optimum where separation and sound quality are near to their individual maxima.

This performance is starkly superior to performance reported for the same test audio using a time-domain convolutional deep transform (CDT) with probabilistic re-synthesis approach [8]. This is not surprising when considering the short-time Fourier transform as a contributory stage of abstraction in a deep architecture. Essentially, the STFT we employ constitutes a layer of abstraction featuring both a filter and a demodulation stage [8], [9], and the inverse STFT constitutes a further filter and synthesis stage. Hence, although the present DNN only features 3 layers, it might be interpreted as featuring a depth of 5 layers if we include those stages of demodulation and synthesis of the STFT and the inverse STFT. Thus, it is not surprising that this network performs far better than the respective 3-layer time domain approach. In addition, this model is more constrained, hence is easier to train than the respective autoencoder [8] and, we note in passing, that (unlike the autoencoder based approach reported previously) the present model showed no sign of bias towards either the male or female voice.

The performance reported here also appears superior to previous methods based on non-negative matrix factorization (NMF) which incorporated deep neural networks as part of the NMF pipeline [4], [5]. While these previous results are not directly comparable with the present results, the ideal binary mask reference allows some comparison to be made; The previous NMF-based models did not reach as close to the ideal binary mask performance as the present model. The advantage of the present approach is likely due to a combination of 1) a relatively large scale network (larger than those reported in [4], [5]) and 2) the probabilistic convolution featured here. The advantage of scale is enhanced by the fact that the present audio data was decimated to a sample rate of 4 kHz, further increasing the effective advantage of scale. It may also be the case that this relation of scale to sampling rate accounts for some performance gains in terms of mitigated aliasing [10].

More generally, given that DNN are inspirsed by, and modeled upon, the neural circuits and function of the brain, it may be that some aspect of the present study offers insight into the possible neural signal processing that might be employed in the human auditory system during cocktail party listening. In principle, a system equivalent or similar to the probabilistic binary mask described here might be implemented in the auditory brain. Indeed, the two minutes of training data applied here is not far in scale from the learning

(adaptation) rate demonstrated in the human auditory perceptual system [11].

ACKNOWLEDGMENT

AJRS was supported by grant EP/L027119/1 from the UK Engineering and Physical Sciences Research Council (EPSRC).